\documentclass[twocolumn,trackchanges]{aastex62}
\usepackage{natbib}
\usepackage{multirow}
\usepackage{graphicx}
\usepackage{tabularx}
\usepackage{verbatim}
\usepackage{color}
\usepackage{ulem}
\usepackage{subfigure}

\graphicspath{{./}{figures/}}

\submitjournal{ApJ}

\shorttitle{Cosmological simulations of satellites around isolated dwarf galaxies}
\shortauthors{Chun et al.}

\begin{document}

\title{Cosmological Simulations of Satellites around Isolated Dwarf Galaxies}

\correspondingauthor{Sungsoo S. Kim}
\email{kwchun@khu.ac.kr, sungsoo.kim@khu.ac.kr}

\author{Kyungwon Chun}
\affil{School of Space Research, Kyung Hee University, 1732 Deogyeong-daero, Yongin-si, Gyeonggi-do 17104, Korea}

\author{Jihye Shin}
\affiliation{Korea Astronomy and Space Science Institute (KASI), 776 Daedeokdae-ro, Yuseong-gu, Daejeon 34055, Korea}

\author{Rory Smith}
\affiliation{Korea Astronomy and Space Science Institute (KASI), 776 Daedeokdae-ro, Yuseong-gu, Daejeon 34055, Korea}

\author{Sungsoo S. Kim}
\affiliation{School of Space Research, Kyung Hee University, 1732 Deogyeong-daero, Yongin-si, Gyeonggi-do 17104, Korea}
\affiliation{Department of Astronomy \& Space Science, Kyung Hee University, 1732 Deogyeong-daero, Yongin-si, Gyeonggi-do 17104, Korea}

\begin{abstract}

We trace the cosmological origin of satellites around a dwarf galaxy using a very high resolution (12~pc$/h$) cosmological hydrodynamic zoom simulation. 
To realistically describe the formation and evolution of small-mass stellar satellites, our model includes a full baryonic physics treatment including a recipe for UV self-shielding. 
We find that the mini-halos form objects resembling dwarf galaxies. 
Despite our high resolution, none of our `mini-halos' form objects that might be considered globular cluster-like. Instead such objects are formed in the host galaxy's gaseous disk.
We investigate why some form stars more efficiently than others. 
We find that the majority of their star forming gas is accreted after reionisation, thus the survival of a mini-halo’s gas to reionisation is not an important factor. Instead, the key factor seems to be the ability for a mini-halo to cool its recently accreted gas, which is more efficient in more massive halos. We find halos in denser environments suffer more mergers, enabling them to grow their mass such that cooling of accreted gas can occur efficiently.
Although the host galaxy is only a dwarf galaxy itself, we find that ram pressure is an efficient means by which accreted mini-halos lose their gas content, both by interacting with hot halo gas but also in direct collisions with the gas disk of the host. The satellites are also disrupted by the tidal forces near the center of the host galaxy.
Compared to the disrupted satellites, surviving satellites are relatively more massive, but tend to infall later into the host galaxy, thus reducing the time they are subjected to destructive environmental mechanisms and dynamical friction. 
In summary, our results suggest that the characteristics of satellites are mainly determined by their ability to efficiently cool gas that is accreted in the redshift range z=3-5, prior to their infall into the host galaxy.

\end{abstract}


\bigskip

\section{Introduction}

In the current Concordance cosmology of Lambda cold dark matter ($\Lambda$CDM), structures are hierarchically built up by merging of smaller structures. 
A galaxy itself is a result of the hierarchical clustering of many smaller buillding blocks. Many of the smaller objects suffer tidal stripping and dynamical friction, and end up spiralling into the central regions of the galaxy.
Although most objects are disrupted, some of these objects, such as globular clusters (GCs), ultra-compact dwarfs (UCDs), dwarf spheroidals (dSphs), ultra-faint dwarfs (UFDs), and dwarf galaxies, can survive as satellites \citep{Ibata1994,Taylor2004,Mashchenko2005b,Saitoh2006,Trenti2015,Gonzalez2016}.

The progenitors of satellites typically form in their own individual low mass dark matter (DM) halos and later fall into the host galaxy \citep{Peebles1984,Mayer2001,Kravtsov2004,Mashchenko2005a,Mashchenko2005b,Penarrubia2008,Lokas2012,Kirby2013a,Trenti2015}.
However, they are believed to form at different epochs with different mass, size, DM content, and star formation history and to have experienced different environmental histories \citep{Mateo1998,McConnachie2006,Evstigneeva2007,Simon2007,Mieske2008}.

Recent studies on the spatial distribution of satellites around a host galaxy reported that the satellites tend to align along filamentary large-scale structures that are connected with the host galaxy \citep{Knebe2004,Gonzalez2013,Tempel2015}. 
Since the satellites form not only in the individual DM halos, but also along the large-scale structures connected to the host galaxy, tracing the formation of the satellites requires us to include a wide range of dynamical scales and cosmological density fluctuations.

It is often believed that reionization or thermal energy feedback will have a high impact on the gas content of small halos, ionising or ejecting their gas, and suppressing their star formation \citep{Shen2014,Sawala2016}. However, efficient radiative cooling can allow some small halos to possess gas which is self-shielded from the ultra violet (UV) radiation \citep{Bland-Hawthorn2015,Wheeler2015,Jethwa2018}. These halos may be very important for the formation of dwarf galaxies and their small satellites.

In this paper, we aim to understand how satellites form around a dwarf galaxy in a cosmological context. 
For this, we perform three cosmological hydrodynamic zoom simulations that form a galaxy whose total mass, $M_{t}$, is $\sim$ 10$^{10}$ M$_{\odot}/h$ at $z=0$. 
We then trace their mini-halos, whose baryon mass ($M_b$) is more than 10$^4$ M$_{\odot}/h$, that form outside the dwarf galaxy, but later fall into the galaxy. 
Hereafter, we refer to the dwarf galaxy as a host galaxy, while the individual mini DM halo that forms outside but later fall into the host galaxy are referred to as `mini-halos'. 
After infall into the host galaxy, mini-halos are referred to as satellites.

This paper is organized as follows. In $\S$\ref{sec:simulations}, we introduce our simulation code and simulation parameters. 
$\S$\ref{sec:hosts} is devoted to describing our three different host galaxies. 
The properties of mini-halos and satellites around the galaxy are described in $\S$\ref{sec:satellites}. 
We compare properties of the surviving satellites with observations in $\S$\ref{sec:observations} and summarize and discuss our results in $\S$\ref{sec:summary}.

\section{Cosmological hydrodynamic simulations}
\label{sec:simulations}
For our cosmological hydrodynamic zoom simulations, we use a parallel N-body/smoothed particle hydrodynamics (SPH) code, Gadget-3 \citep{Springel2005}. 
To realistically reproduce small-scale gaseous structures, one needs to include low temperature cooling below 10$^4$ K \citep{Saitoh2006}. 
For this, we include the following processes; 1) radiative cooling/heating down to 10 K \citep{Ferland1998}, 2) star formation from gas clouds that satisfy: $n_H \geq$ 100 cm$^{-3}$, T $<$ 10$^4$ K, and $\nabla \cdot v <$ 0 \citep{Saitoh2008}, and 3) thermal energy feedback by supernova (SN) explosions \citep{Okamoto2008}. 
Uniform ultra-violet (UV) radiation by cosmic reionization at $z\sim8.9$ is considered in the radiative cooling/heating terms \citep{Haardt1996}. 
Critically, high-density gas with $n_H \geq$ 0.014 cm$^{-3}$ is assumed to be self-shielded from the UV radiation \citep{Tajiri1998,Sawala2010}. 
The full details of this scheme are described in \citet{Shin2014} and references therein. 

\citet{Saitoh2009} reported that SPH simulations with individual time-stepping schemes do not properly describe the effects of SN feedback because gas particles surrounding an exploding star particle cannot react instantly to the SN event. 
This leads to very strong feedback, larger than the original SN energy, and thus violates energy conservation. 
To avoid this, we implemented a time-step update algorithm introduced by \citet{Durier2012}, which is a modified version of the time-step limiter of \citet{Saitoh2009}. 

We used the post-Planck cosmological model of $\Omega_{m} = 0.3,~\Omega_{\Lambda} = 0.7,~\Omega_{b} = 0.048$, and $h = 0.68$ \citep{Planck2014}, and the power spectrum under this cosmology is generated by the CAMB package\footnote{http://camb.info/}  \citep{Lewis2000}. 
A linear approximation for the structure growth is imposed until $z=49$. We start non-linear N-body/SPH calculations from this redshift onwards. 
Initial conditions for particle positions and velocities are generated by a MUSIC package \citep{Hahn2011}. 

To efficiently consider both the large-scale and small-scale structures that enable a dwarf galaxy and satellites to form, respectively, we perform simulations with two different resolutions, i.e., zoom simulations. 
First, a DM-only simulation of an (8~Mpc$/h$)$^3$ uniform box is performed with $\sim$~17 million low-resolution particles.
We then search for DM halo candidates for our host galaxies, that have an M$_{t}$ of ∼10$^{10}$~M$_{\odot}/h$ at $z=0$ and are located in isolated environments\footnote{An isolated environment for a halo is defined as having no other halos whose mass is more than 50~\% of the halo within a radius 5 times larger than $r_{vir}$ of the halo.}.
The lagrangian volume for the dwarf galaxy-like halo is traced back to the initial conditions, and more detailed and high resolution density fluctuations than those of the DM-only simulation are added within this volume (i.e. the zoom region).
Thus, the zoom region is composed of high-resolution gas and DM particles, while the rest of the simulation box consists of low-resolution DM particles. 
Next, we perform multi-resolution hydrodynamic simulations with $\sim$~51 million particles, whose high-resolution particle mass for DM and gas is $\sim$ 4$\times$10$^3$~M$_{\odot}/h$ and $\sim$ 8$\times$10$^2$~M$_{\odot}/h$, respectively.
The initial mass of a star particle is one third that of the original gas particle mass. Each star particle contains a stellar population that follows the initial mass function of \cite{Kroupa2001} and suffers mass loss following the stellar evolution model of \cite{Hurley2000}.
To avoid contamination by low-resolution particles in the high-resolution region, we increase the Lagrangian volume to cover 5 times the virial radius $r_{vir}$ of the dwarf galaxy-like halo \citep{Onorbe2014}. 
Gravitational softening length for the high-resolution region is fixed as 130~pc$/h$ on a comoving scale for $z>10$ and 12~pc$/h$ on a physical scale for $z\leq10$. For better statistical results, we performed three different zoom simulations using different galaxy targets until $z=0$. 

Halo and sub-halo structures are identified with the halo finding algorithm, Amiga Halo Finder (AHF), which groups gravitationally-bound particles centered on a local density peak \citep{Knollmann2009}. 
We define the virial radius of the halos as the radius that the average density within it is 200 times the cosmological background density.
In AHF, the criteria to determine if a particle is bound to a halo is based on the particle's relative velocity and position with respect to the halo. The gas thermal energy is also considered for gas particles.

\begin{figure*}[h]
\centering
\includegraphics[width=0.88\textwidth, height=0.88\textheight]{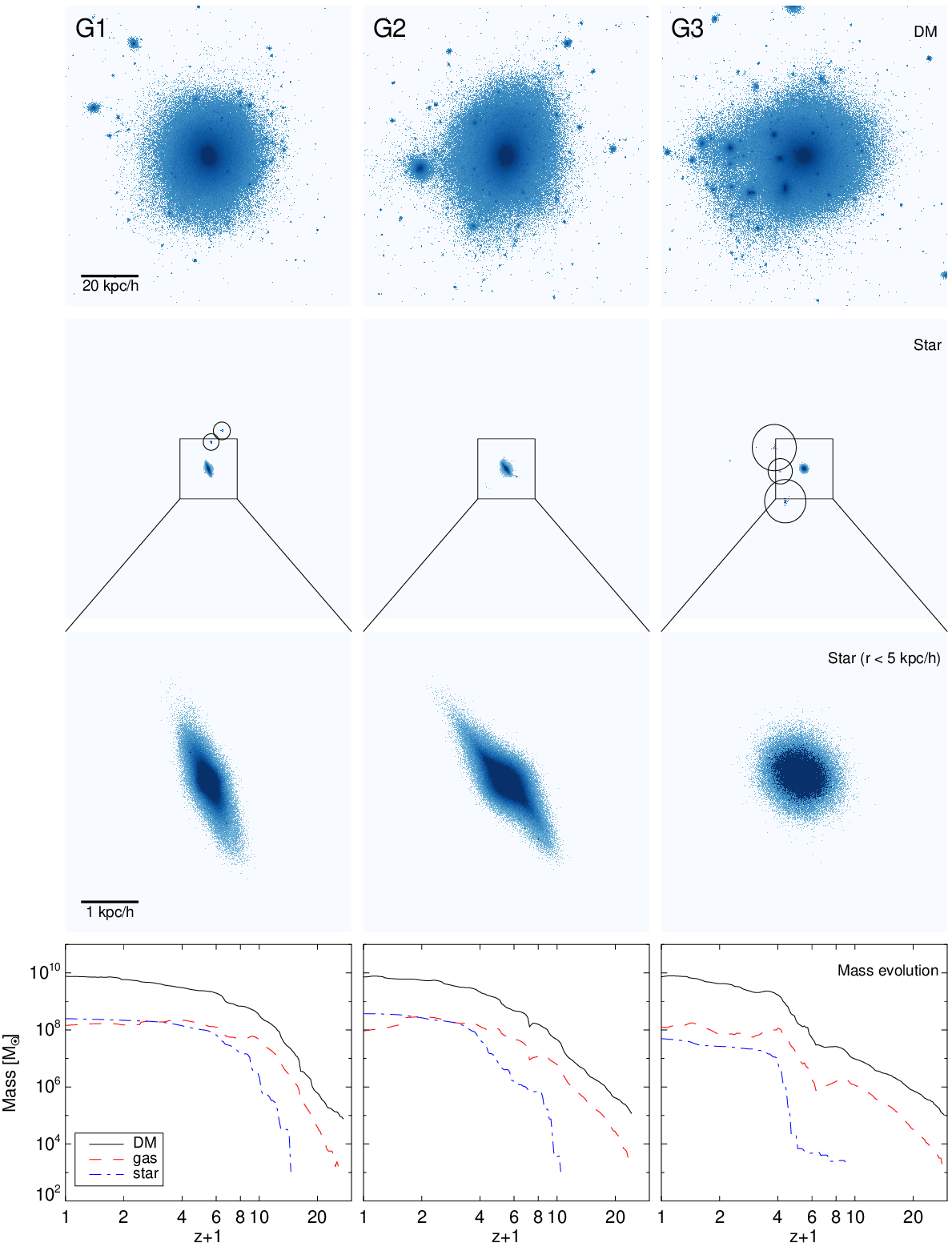}
\caption{Structures and mass growth history of G1, G2, and G3. The first and second rows show structures of DM and stellar components inside the virial radius at $z=0$. The black circles in the second row indicate the virial radius of surviving satellites containing stars. The third row shows the stellar structure at the central 5 kpc$/h$ of the galaxies. The bottom row shows the mass growth history of host galaxies until $z=0$. Solid black, dashed red, and dashed-dot blue lines represent the evolutions of DM, gas, and star contents, respectively.}
\label{fig:structure}
\end{figure*}

\begin{table*}
	\centering
	\caption{Properties of the host galaxies. Columns 3-7 are the quantities at $z=0$: total (DM+gas+star), baryon (gas+star), stellar mass, virial radius, and maximum rotation velocity, respectively. Columns 9-10 describe the total number of satellite halos that have fallen into a host galaxy and surviving satellites at $z=0$, respectively. Columns 10-11 are epochs of galaxy formation and last major merger (LMM).}
	\label{tab:table1}
	\begin{tabular}{cccccccccccccccccccccc}
		\hline 
        Run & Type & $M_t$ & \multicolumn{1}{c}{$M_b$} & $M_s$ & \multicolumn{1}{c}{$r_{vir}$} & 
        \multicolumn{1}{c}{$V_{max}$} & {$N_{sat}$} & $N_{sat}(z=0)$ & $z_{form}$ & $z_{LMM}$ \\[1.0ex]
        \cline{3-5} \\[-3.0ex]
 & & & $[10^{9}$M$_{\odot}]$ & & [kpc$/h$] & [km/s] & & &\\
 		\hline
        G1 & Disk & 7.95 & 0.391 & 0.247 & 32.46 & 51.06 & 48 & 3 & 26.3 & 5.98 \\
        G2 & Disk & 7.71 & 0.463 & 0.372 & 32.12 & 64.73 & 34 & 2 & 23.2 & 6.21 \\
        G3 & Elliptical & 7.46 & 0.178 & 0.050 & 31.78 & 36.24 & 15 & 3 & 27.9 & 3.25 \\
        \hline
	\end{tabular}
\end{table*}

\section{Properties of host galaxies}
\label{sec:hosts}

We name the host galaxy of the zoom simulations G1, G2, and G3 (see Table \ref{tab:table1} and Fig. \ref{fig:structure}). 
G1 and G2 are similar in that they have a disk morphology of gas and stars. Their disks first appear at $z~\sim$ 5.8 and 3.3, and they have a maximum circular velocity of 51~km/s and 65~km/s, respectively. 
G3 shows a more spherical morphology, with a maximum circular velocity 36 km/s. 
The rotational velocity of G1 and G2 successfully reproduce those of observed dwarf spirals with comparable mass, e.g. 42-153~km/s \citep{Swaters2009}. 
The morphological difference between G3 and the other two appears to be a result of different epoch of the last major merger (LMM)\footnote{We define a major merging to be the mass ratio between galaxies smaller than 1:3.}, $z_{LMM}$; both G1 and G2 undergo LMM at $z_{LMM}\sim6$, while G3 at $z_{LMM}\sim3.3$. The disk structure of G1 and G2 lasts until the end of the run, $z=0$, since there are no other merging events sufficient to disturb the disk structure. 

Fig \ref{fig:structure} shows DM and/or stellar structures around the host galaxies of G1, G2, and G3 at $z=0$ (first and second rows). 
Black circles in the second row indicate the surviving satellites that are originally from outside of the host galaxies. 
Surrounding the host galaxies, there are more surviving satellites, however most of them do not contain any stars. 
Thus, the stellar satellites shown in the second panel are much rarer than the satellites shown in the first panel.  
G1 and G2 are clearly more disk-like than G3 (third row). 

Bottom row of Fig.\ref{fig:structure} shows the mass evolution of G1, G2 and G3. The DM mass of G1 and G2 grow relatively faster than that of G3; half of the final mass is accumulated at $z\sim$ 2.4, 1.9, and 1.3, for G1, G2, and G3, respectively. 
We confirm the mass growth history of G1, G2, and G3 with that of all the halos ($\sim10^{10}$~M$_{\odot}/h$ at $z=0$) in the entire box of our low resolution DM-only simulations. We find that they have quite typical mass growth histories -- their mass growth history is within the 1$\sigma$ range of the variety of mass growth histories shown by all the halos.
Accumulated gas at the DM halo center starts to form stars at $z=13.6$, 9.5, and 7.9 for G1, G2, and G3, respectively. 

The stellar mass of G1 and G2 rapidly increase after the first star formation, while that of G3 hardly grows until when G3 undergoes a major merger at $z\sim$ 3.25.
Since the gas accretion and star formation of G3 takes place late compared to those of G1 and G2, the ratio of stellar mass to baryon mass at $z=0$, $M_{s}/M_{b}$, of G3 is 0.28, which is lower than those of G1 and G2 at 0.63 and 0.80, respectively. 
Here, the accumulated $M_{s}$ for G1, G2, and G3 is estimated to have V-band absolute magnitudes of $M_{V}$ $\sim$ -15.84, -16.32, and -14.01, respectively. 
This closely reproduces $M_{V}>$~-18 of observed dwarf galaxies \citep{Schombert1995,Mateo1998}, where $M_{V}$ in the simulations is calculated with the $M_{s}-M_{V}$ relation, $\log_{10}(\frac{M_s}{\textrm{M}_{\odot}})=2.37-0.38M_V$, of given by \citet{Brooks2013}. 
As SN produce metals and more massive galaxies are better able to hold onto these metals, G1 and G2, that grow relatively faster, end up with larger [Fe/H] (-0.61 and -0.59 respectively), while G3 has [Fe/H] = -1.03. These results are a good match to the observed stellar mass-stellar metallicity relation of \citet{Kirby2013b}.
In addition, it is possible that the accretion of pristine gas could also play a role in regulating the metal abundances of the host galaxies.

Thanks to the realistic baryonic treatment described in Section \ref{sec:simulations} as well as the high mass resolution, the three host galaxies well reproduce the observed properties of dwarf galaxies, as described above.
\cite{Shin2014} showed that the same choice of recipes for the baryonic physics also well reproduce the observed star formation rates in significantly more massive systems such as a Milky Way-like model disk galaxy, and also the star formation history of the universe in cosmological simulations. 
We will later show in Section \ref{sec:observations} that our model is also reliable for modelling even lower mass systems than our host galaxies by comparing the properties of the host galaxy's satellites to observed dwarfs of similar luminosity.

G1, G2, and G3 accrete a total of 97 mini-halos until $z=0$, but only 8 of them survive until the end of the run. 
Among them, only 5 satellites include stars at $z=0$ (black circles of Fig.\ref{fig:structure}.)

The host galaxies grow their mass from merging events with other smaller halos and smooth accretion. 
We find that $\sim50~\%$ ($\sim42~\%$) of the DM (gas) mass of the host galaxies comes from merging with other halos.
The rest $\sim50~\%$ ($\sim48~\%$) of the DM (gas) mass growth comes from smooth accretion or unresolved halos below our mass resolution limit ($\sim$ 7$\times$10$^4$~M$_{\odot}/h$ for halos with less than 20 DM particles).

\section{Properties of satellites}
\label{sec:satellites}

\subsection{Pre-infall}

\begin{figure}
\centering
\includegraphics[width=0.45\textwidth]{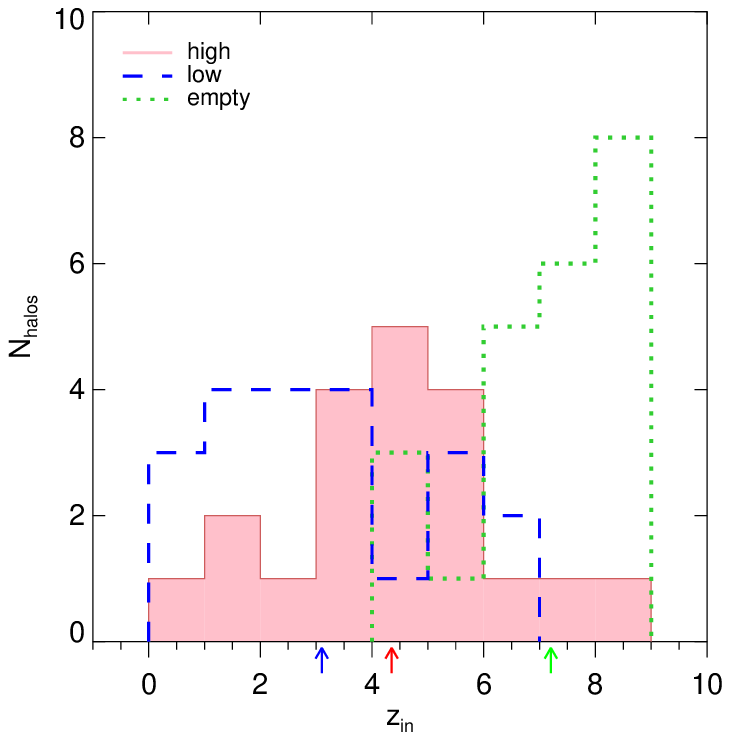}
\caption{Histogram of redshift that the mini-halos infall to the host galaxy. Red, blue, and green histograms indicate high stellar fraction, low stellar fraction, and empty halos, respectively. The arrows below the x-axis show the mean values of $z_{in}$ of each subsamples.}
\label{fig:zin}
\end{figure}
 
Observed galaxies often contain a wealth of different types of satellites: e.g. GCs, dSphs, UFDs, and dwarf galaxies. The properties of their satellites have been much better constrained through recent high-resolution photometries and spectroscopies \citep{Mateo1998,McConnachie2006,Evstigneeva2007,Simon2007,Mieske2008}. Despite this, their formation paths remain unclear. 
To better understand the evolution of the satellites in our simulations, we analyze the properties of the 64 mini-halos that form outside the host galaxy, before cosmic reionization, but fall into the host galaxy after reionization, from a total of 97 mini-halos that fall into the three host dwarf galaxies. 
We neglect 33 of the 97 mini-halos because they fall into the host galaxy before reionization, but we confirm that this choice does not impact on our main conclusions.

First, we classify the 64 mini-halos into three subsamples, high stellar fraction, low stellar fraction, and empty, according to the stellar mass fraction, $M_s/M_t$, when the mini-halos fall into the host galaxy, $z_{in}$. Our choice of $M_s/M_t (z_{in})$ limit roughly divides the subsample of halos containing stars in half.
Twenty halos fall in the high stellar fraction subsample, with $M_s/M_t (z_{in}) > 3\times10^{-4}$. The low stellar fraction subsample, with $M_s/M_t (z_{in}) < 3\times10^{-4}$ contains twenty-one halos. Twenty-three mini-halos do not form any stars at all and are placed in the empty subsample. 

Fig. \ref{fig:zin} shows the infall redshift ($z_{in}$) histogram of the mini-halos in the three subsamples. 
We can see that the empty halos (green histogram) fall into the host galaxy earlier than in the other halo subsamples. 
Indeed, the mean value of the infall redshift of the empty halos is $z_{in}=7.2$ which is higher than that of the high  stellar fraction (red histogram; $z_{in}=4.35$) and low stellar fraction (blue histogram; $z_{in}=3.1$) subsamples.
It is likely that the very early infall of the empty halos does not permit sufficient time for these halos to form stars.
On the other hand, the high stellar fraction subsample typically falls in earlier than the low stellar fraction subsample, so we cannot fully explain the differences in efficiency of star formation between all the subsamples simply with their infall time.

\begin{figure}
\centering
\includegraphics[width=0.45\textwidth]{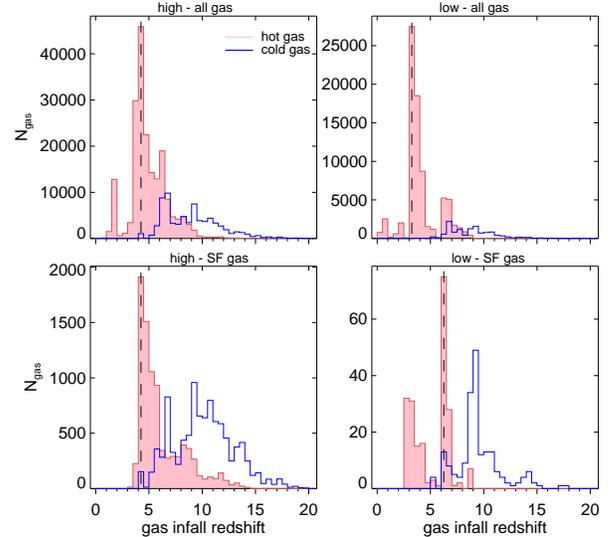}
\caption{Histograms of redshift that the gas particles fall into the mini-halos. Red and blue histograms mean hot gas (T$>$10$^4$~K) and cold gas (T$<$10$^4$~K), respectively. Two left figures show the histograms of all gas (the left upper panel) and star forming gas (the left lower panel) that fall into the high stellar fraction subsample halos. Two right panels are for the low stellar fraction subsample halos. The vertical dashed line in each histograms indicates the peak of gas accretion.}
\label{fig:ginfall}
\end{figure}

The histograms of Fig. \ref{fig:ginfall} show the infall redshift of all the gas (upper panels) and gas particles that will later contribute to star-formation (bottom panels), when they are accreted onto the mini-halos of the high stellar fraction (left panels) and low stellar fraction subsample (right panels). 
In this figure, we consider the contribution of gas from both mergers and smooth accretion. The contribution from mergers is only 15~\% and 90~\% of the merger gas is accreted after cosmic reionization. Although mergers are less significant than smooth accretion in terms of total accreted gas, in fact 70~\% of the cold gas (blue histogram) with T$~<~10^4~$K that is accreted externally after cosmic reionization originates from mergers.
We can see that most cold gas falls into the mini-halos at higher redshift and it is 24.7~\% of the total gas. 

The high and low subsamples do accrete some gas before cosmic reionization, but the amount is only 14.5~\% and 6.2~\% of the total accreted gas, respectively.
We note that the high and low stellar fraction subsamples accrete 40~\% and 30~\% of their gas that later contributes to star formation before reionization, respectively. 
While this is certainly not negligible, the fact that the fraction of the gas and the trend of the gas accretion are quite similar between the subsamples means that differing gas accretion prior to reionisation is unlikely to be the source of the widely differing star formation between the high and low stellar fraction subsamples.

As low density gas is reionized by UV radiation after the epoch of cosmic reionization, hot gas with T$~>~10^4~$K (red histograms) dominates the budget of accreted gas, and is easily the biggest contributor to gas that will eventually be involved in star formation. 
The majority of this hot gas is accreted in the redshift range $z=3-5$, in all the subsamples. 
This also demonstrates that the ability for these mini-halos to manage to hold onto their gas against reionization is not directly related to whether they will end up with high stellar fractions.

\begin{figure}
\centering
\includegraphics[width=0.45\textwidth]{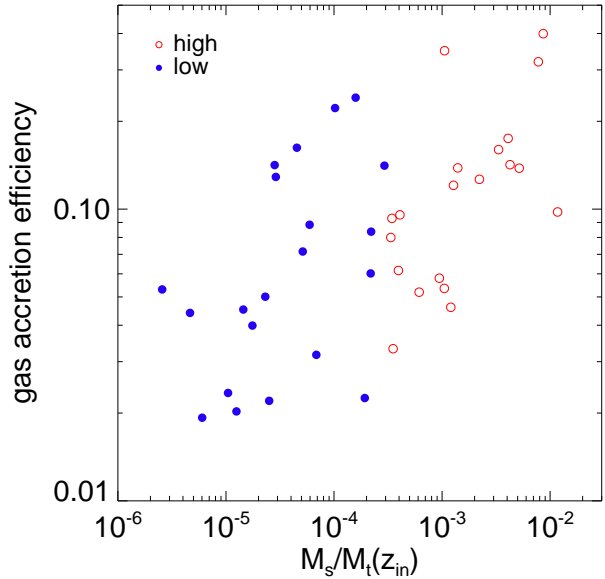}
\caption{Scatter plot of gas accretion efficiency versus $M_s/M_t(z_{in})$, where gas accretion efficiency is defined as total accreted gas mass compared to $M_t(z_{in})$. Open red and filled blue circles indicate high and low stellar fraction subsample halos, respectively.}
\label{fig:efficiency}
\end{figure}

To test for other possible contributing differences between the subsamples, we normalize the total accreted gas mass between $z_{form}$ and $z_{in}$ by the halo mass $M_t(z_{in})$ and refer to this as the `gas accretion efficiency'.
We find there is a broad trend with $M_s/M_t(z_{in})$ (Fig. \ref{fig:efficiency}). 
But the large amount of scatter in the trend means that it is easy to find objects with a broad range of gas accretion efficiency at fixed $M_s/M_t(z_{in})$, which shows that the difference between our high and low stellar fraction subsamples is not just a result of differing gas accretion efficiency.

The gas accretion of both subsamples peaks at $z=3-5$ as we mentioned above, whereas the accretion of hot gas that will later contribute to star formation depends on the subsample.
The vertical dashed lines in Fig. \ref{fig:ginfall} indicate the peak of the total gas or the star forming gas accretion. It shows the dependence on the subsamples directly.
In the case of the high stellar fraction halos, most of the gas that later contributes to star formation is accreted at the same time as the peak of the total gas accretion. 
On the other hand, the low stellar fraction subsample halos accrete most gas contributing to star formation at early redshifts, $z=6-7$ (right panels of Fig. \ref{fig:ginfall}).
This is a clear difference in behaviour between the two subsamples, and it strongly suggests that \textit{the high stellar fraction subsample is able to much more efficiently turn its accreted hot gas into cool star forming gas than the low stellar fraction subsample}.

\begin{figure}
\centering
  \includegraphics[width=0.45\textwidth]{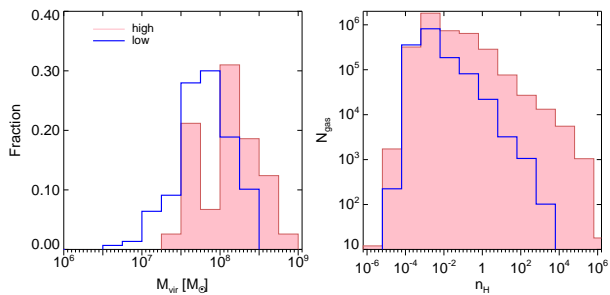}
\caption{Histograms of halo virial mass (left panel) and hydrogen number density, $n_H$, of gas particles in the halos (right panel) at $z=3-5$. The number of halos and gas particles is accumulated using all snapshots between $z=3$ and $z=5$. Red and blue histograms mean high and low stellar fraction subsample halos, respectively.}
\label{fig:mass}
\end{figure}

To try and understand why the two subsamples differ in their ability to cool accreted gas, we consider if their halo masses differ significantly.
In the left panel of Fig. \ref{fig:mass}, the histogram indicates the virial mass of both subsample halos at $z=3-5$. 
The number of halos is combined using all snapshots between $z=3$ and $z=5$. There is a clear tendency for the high stellar fraction subsample to have greater halos masses.
The mean virial mass of the high stellar fraction (red histograms) and the low stellar fraction (blue histograms) subsample halos during the period is $1.79\times10^8$~M$_{\odot}/h$ and $8.32\times10^7$~M$_{\odot}/h$, respectively. The NFW concentration parameter (c$_{NFW}\sim$3.7) of the high stellar fraction subsample halos is also slightly greater than that (c$_{NFW}\sim$3.0) of the low stellar fraction subsample halos at these redshifts.

\cite{Okamoto2008b} found that reionization efficiently suppresses gas accretion in halos with mass below a characteristic mass, $M_c(z)$. They define the characteristic mass as the halo mass when a halo has a baryon mass fraction of half the cosmic mean value in their simulations.
However in our simulations we find that, although the virial mass of both subsamples is smaller than the threshold mass of \cite{Okamoto2008b}, they can continue to accrete gas after reionization (Fig. \ref{fig:mc}). 
Therefore, we can not explain our accretion histories based on their recipe. This may be because our simulations include a baryonic physics recipe for low temperature cooling below $10^4~$K and self-shielding (see Appendix for more details).
We conclude it is likely that the higher mass, higher mass concentration, and thus deeper gravitational potential well of the high stellar fraction subsample halos enable them to compress the accreted gas to higher density and thus more efficiently cool their accreted gas than in the low stellar fraction subsample. 
In addition, it is likely that the gas heated by SNe can remain dense and thus cool itself again more efficiently in the deeper gravitational potential well of the high stellar fraction subsample halos than in the low stellar fraction subsample halos, which is another way to boost their star formation efficiency.

To investigate further, we calculate density $n_H$ of gas inside the halos at $z=3-5$ (right panel of Fig. \ref{fig:mass}). 
The gas density distribution is combined using all snapshots between $z=3$ and $z=5$. 
In this histogram, we can see that the proportion of dense gas in the high stellar fraction subsample halos is larger than that of the low stellar fraction subsample halos. 
The amount of self-shielded gas among the entire gas in high and low stellar fraction subsample halos is 36.8~\% and 14.3~\%, respectively.
The self-shielding enables gas to become cool star forming gas by avoiding the effects of UV heating.
Even if the fraction of the gas satisfying the star formation criteria is only 1~\% among all of the gas in the high stellar fraction subsample halos, this fraction is 20 times higher than that of the low stellar fraction subsample halos.
In addition, in the case of the high stellar fraction subsample halos, they can form dense cloud with $n_H\sim$10$^6~$cm$^{-3}$, while in the low stellar fraction subsample halos, they can only form gas cloud up to $n_H\sim3\times10^3~$cm$^{-3}$. 
In conclusion, the mass of the halo is an important parameter controlling the formation and lifetimes of dense star forming gas clouds whose presence significantly increases the stellar mass fractions of the mini-halos.

\begin{figure}
\centering
\includegraphics[width=0.45\textwidth]{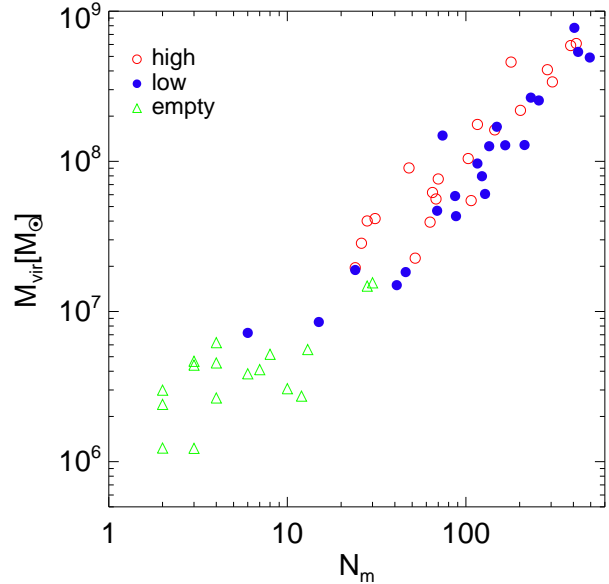}
\caption{Scatter plot of $M_t(z_{in})$ versus $N_m$ for three subsample halos. Open red and filled blue circles indicate high stellar fraction, low stellar fraction, and open green triangles represent empty subsample halos.}
\label{fig:merge}
\end{figure}

As the mini-halos merge with the smaller structures and/or accrete matter, their mass increases. 
For more detailed understanding of how the mini-halos increase in mass, we looked into how $M_t(z_{in})$ is correlated with $N_m$, where $N_m$ is the number of merging events\footnote{The number of merging events is counted when two or more individual halos identified by AHF at the i time step are identified as a single halo at the i + 1 time step. Mass increase that is not attributed to halo merging is defined as smooth mass accretion.} until $z_{in}$ (Fig. \ref{fig:merge}). 
We can see that the halos with larger $N_m$ grow their mass more.
The empty halos undergo few merging events because of their early infall. 
On the other hand, the high and low stellar fraction subsample halos have larger $N_m$ than the empty halos. 
Although both halos have a similar range of $N_m$, the high stellar fraction subsample halos undergo more frequent merging events because they fall into the host galaxy earlier than the low stellar fraction subsample halos. 
\cite{Maccio2017} looked for a strong burst in the star formation after major merger events, by providing an additional source of gas or tidally compressing the existing gas, using a simple visual inspection (see Fig. 14 of their paper). We also investigated whether merger events might contribute to the star formation in a similar manner as in \cite{Maccio2017}. However, we find no clear evidence for a strong connection between the merger events and the star formation activity of the galaxies involved.

Although there is a strong correlations between $M_t(z_{in})-N_m$, we confirm that only 5 out of the 64 mini-halos increase their mass more by merging than smooth accretion. 
Therefore it is clear that $N_{m}$ can be considered a proxy for total mass growth (including both smooth accretion and mergers).  We infer that $N_{m}$ indicates the environmental history of a mini-halo; e.g., mini-halos with larger $N_{m}$ have likely passed through denser regions, where more matter was added by both smooth accretion and merging events \citep{Fakhouri2009}.

We also investigated whether the formation mass of the halos contributes to their mass growth or not, but we found no clear correlation.

\begin{figure}
\includegraphics[width=0.45\textwidth]{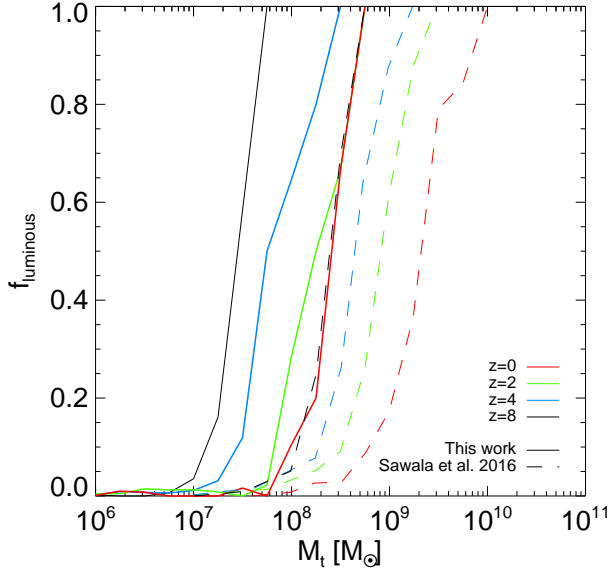}
\caption{Fraction of halos having luminous (star) particles as a function of the total mass. The solid and dashed lines represent results of this work and \cite{Sawala2016}, respectively. The red, green, blue, and black colors indicate different redshift of $z=$ 0, 2, 4, and 8.}
\label{fig:Sawala}
\end{figure}

Contrary to a recent result of hydrodynamic simulations showing that halos of $M_t<10^9$~M$_{\odot}/h$ rarely contain any stars \citep{Sawala2016}, our results show that all halos of $M_t>6\times10^8$~M$_{\odot}/h$ contain more than $M_s>10^4$~M$_{\odot}/h$ of stars, where $10^4$~M$_{\odot}/h$ corresponds to the gas particle mass in the highest resolution run of \cite{Sawala2016} (see Fig. \ref{fig:Sawala}). 
We infer that this difference stems from a fact that the following two parameters are applied differently in the two simulations: the epoch of reionization (z=11.5 in theirs, while z=8.9 in ours) and the existence of self-shielding (not included in theirs, while included in ours). 
Due to the lower reionization epoch and the self-shielding used in our simulations, gas clouds can reside deep in the central regions of the DM halo, avoiding the effects of UV heating, and thus form stars even in the low-mass regime. 
Indeed, these low-mass objects of M$_t\sim~6\times10^8$~M$_{\odot}/h$ are reported to be candidate sites to form the UFD \citep{Wheeler2015}. 

Furthermore, our simulations trace the isolated dwarf galaxy and their satellites, but \citet{Sawala2016} focused on the Local Group environment.
Therefore, there is a possibility that the environmental differences between the two simulations could also contribute to the mass difference of the halos containing stars.

One clear result is that reionisation is unable to empty all of the gas out of halos, even when they have quite low masses ($5.8\times10^5$~M$_{\odot}/h~<~M_t~<~1.5\times10^8$~M$_{\odot}/h$). 
This is because self-shielding plays an important role in counteracting the effects of reionisation. 
For example, our low stellar fraction halos lose only 13~\% of their gas mass due to reionisation. 
In fact, the high stellar fraction sample loses negligible gas. 
Furthermore, most of the gas that will contribute to star formation is accreted later, after reionization. 
However this does not mean that reionization is insignificant in all ways for mini-halos. 
Reionisation is responsible for causing the accreted gas to be hot (T$~>~10^4~$K). 
Thus, when it enters a mini-halo, the halo's gravitational potential well must be sufficient to compress the accreted gas to high enough density to enable more efficient cooling, and star formation. 
Indeed, the self-shielding continues to aid this process even long after reionisation has finished, by enabling gas that has been heated by stellar feedback to cool back down more efficiently, despite heating from the background UV field of the Universe.

\subsection{Post-infall}

\begin{figure*}
\begin{center}
\includegraphics[width=0.9\textwidth]{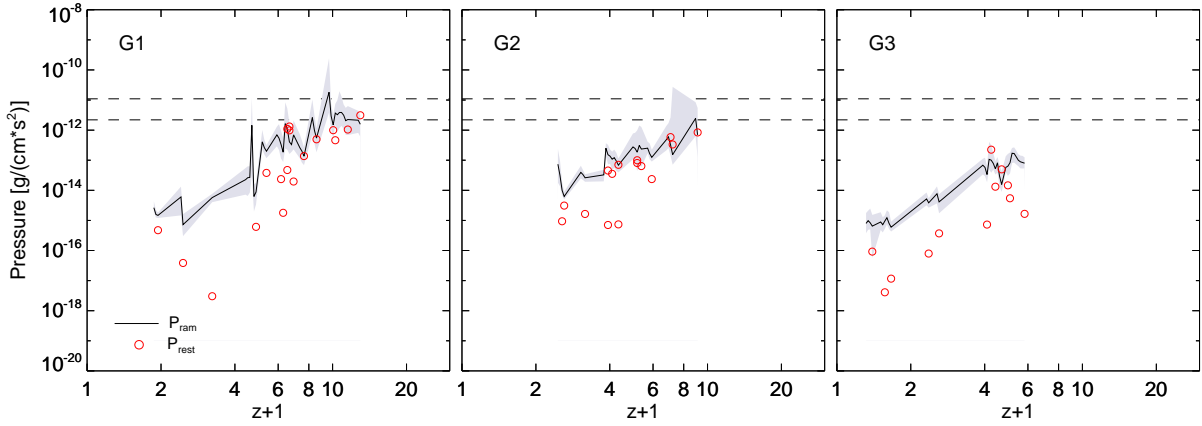}
\end{center}
\caption{Evolution of the ram pressure of G1, G2, and G3. The black solid lines are the median values of ram pressures felt by all satellites passing though the outer halo (r$>0.5r_{vir}$) at that time, while the shaded regions indicates the first and third quartiles of the ram pressures. The red circles indicate the restoring force per unit area of the infalling satellites. For reference, the range of peak ram pressures experienced by satellites of the Virgo cluster, 1,000 and 5,000~cm$^{-3}$~(km/s)$^2$, are shown by black dashed horizontal lines \citep{Vollmer2001}.}
\label{fig:ram_pressure2}
\end{figure*}

\begin{figure*}
\begin{center}
\includegraphics[width=0.8\textwidth]{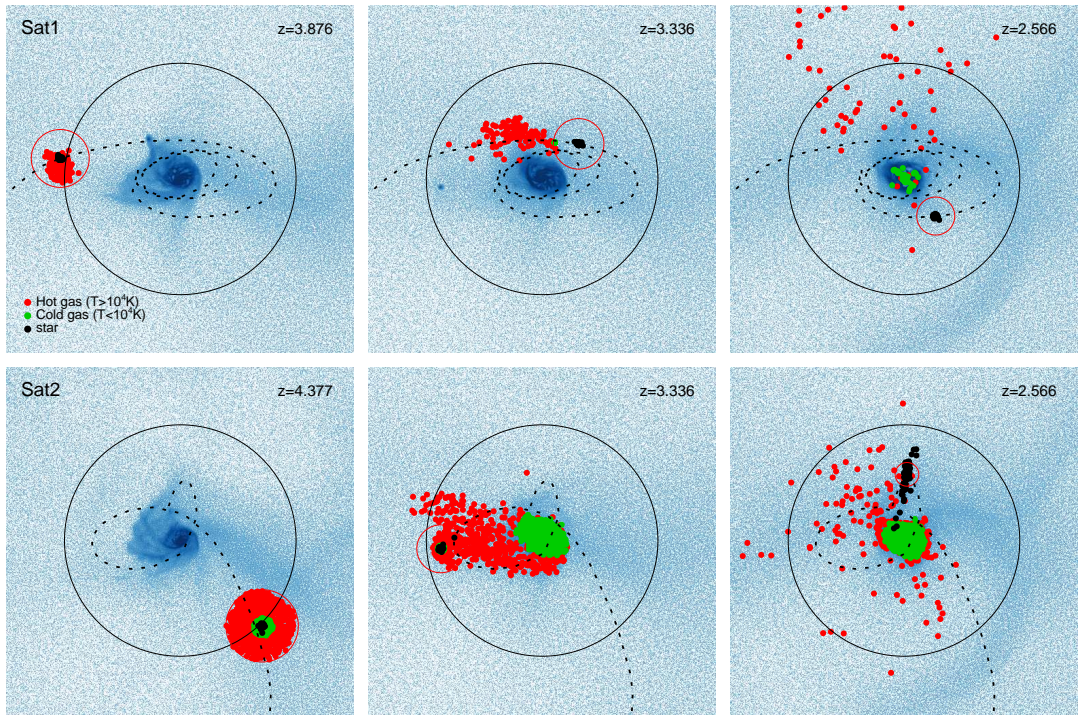}
\end{center}
\caption{Orbital trajectories and gas distribution of two representative satellites of G1: Sat1 (upper panels) and Sat2 (lower panels). The red, green, and black dots indicate the hot gas (T$>$10$^4$~K), cold gas (T$<$10$^4$~K) and star particles, respectively. The black and red circles indicate the virial radius of G1 and the satellites, respectively. The black dashed lines show the orbital trajectory of the satellites. Projected gas density of G1 is overplotted in the background with a blue color scheme.}
\label{fig:ram_pressure}
\end{figure*}

Ram pressure stripping is often considered to be an environmental mechanism predominantly arising in the cluster environments. However, we find that the mini-halos which fall into their host galaxy, so-called satellites, lose a significant portion of their gas content due to ram pressure stripping. 
The ram pressure that host galaxies subject their satellites to, $P_{ram}$, is calculated by
\begin{equation}
P_{ram} = \rho_{gas}v_{gas}^2,
\end{equation}
where $\rho_{gas}$ is the density of the gas through which the satellite passes and v$_{gas}$ is the relative velocity of the surrounding gas with respect to the satellite. The properties of the surrounding gas are based on the average density and average relative velocity of all gas particles within $r_{sat}<r<1.5~r_{sat}$, where $r_{sat}$ is the virial radius of the satellite.

We measure the ram pressure experienced by each satellite found in the outer halo ($r > 0.5~r_{vir}$). The black line in Fig. \ref{fig:ram_pressure2} shows the median value of the ram pressure for all the satellites in the outer halo at that moment in time. The shaded regions show the first and third quartiles of the ram pressure, which gives some sense of the range of ram pressure strengths experienced between individual satellites. We see a clear trend for a decrease in the ram pressures with time, which occurs because the outer halo's gas density is continuously decreasing with time.

For comparison to the ram pressures, the red circles indicate the gravitational restoring force per unit area of the infalling satellites that contain gas.
Here, the restoring force per unit area of the satellites, $P_{res}$, is computed using 
\begin{equation}
P_{res}=2\frac{GM_{tot,sat}(r)\rho_{gas,sat}(r)}{r},
\end{equation}
where $M_{tot,sat}(r)$ and $\rho_{gas,sat}(r)$ are total mass and gas density of the satellite within a radius $r$ \citep{McCarthy2008}.
We measure restoring forces at the virial radius of the satellite, in order to see when ram pressure is sufficient to strip their hot halo gas which otherwise would later cool.

We find that many of infalling satellites have a smaller restoring force per unit area compared to the ram pressure of the host galaxy, and thus lose significant gas contents due to ram pressure stripping from the hot gas in the outer regions of the host halo ($r>0.5~r_{vir}$). 
We find that 54 of the 97 satellites lose their entire gas contents during the first few orbital passages, while the others lose $\sim$ 46~\% on average during their life time. 
Among the 54 satellites, 35 satellites lose their entire gas content when the ram pressure is most effective. 
Among the 35 satellites, 17 satellites lose a negligible DM mass ($\sim 5~\%$), while the others lose $\sim$ 55~\% before their gas disappears. 
Therefore, we infer that the former 17 satellites lose their entire gas by ram pressure stripping, while the latter 18 satellites suffer both ram pressure stripping and gravitational tidal stripping simultaneously.

In Fig. \ref{fig:ram_pressure}, the orbital trajectories of two representative satellites, Sat1 and Sat2, are shown together with their gas distribution.
The orbital trajectories are traced as relative position of the satellites with respect to the center of the host galaxy, and we track their positions by following the main progenitor of the merger tree.
These two satellites lose their dense cold gas (filled green symbols) as well as hot gas (filled red symbols) during the first few orbital passages. 
In the figure, the colour of a symbol denotes the current temperature at the moment of the snapshot (as opposed to the temperature at the infall time into the host). This means that the green points seen near the host centre (cold gas) may originally have been hot gas denoted by red points in a previous snapshot that has since cooled onto the host's disk.

Sat1 has $9.1\times10^{5}$~M$_{\odot}/h$ of gas content when it enters the host galaxy, $z_{in}=3.876$. 
This satellite loses all of its gas content in the hot gas halo of the host galaxy at $z=3.51$, before it reaches the centre of the host galaxy.
As Sat1 loses only 10~\% of DM content by the time its gas is stripped, we can conclude that Sat1 loses the gas content due to the ram pressure stripping from the hot gas in the host galaxy's outer halo.
While Sat1 loses its gas content at early times due to the ram pressure, the stellar mass loss of Sat1 is delayed until $z=1.82$ when 77~\% of the outer DM halo has been stripped. 

Sat2 has $4.72\times10^{6}$~M$_{\odot}/h$ of gas when it falls into the host galaxy at $z_{in}=4.377$.
While Sat1 reaches the first pericenter after 1.5~Gyr from $z_{in}$, Sat2 arrives at first pericenter only 200~Myr after $z_{in}$. 
Since Sat2 quickly suffers strong tidal forces near the host galaxy's center, Sat2 rapidly loses its DM content.
It loses half of its DM as well as some gas content prior to reaching first pericenter at $z=3.97$.
It then crosses the host galaxy's disk, and suddenly loses 96~\% of its gas content and 75~\% of its DM mass at infall.
After second pericentre passage at $z=2.86$, it loses the remaining gas, and also some of its stars are stripped for the first time. At this point in time it has lost 89~\% of its DM mass at infall.
Thus, we conclude that most of the gas contents of Sat2 is stripped by a combination of ram pressure from both the hot diffuse gas of the host's gas halo and the cold dense gas of the host's disk. In comparison, the gas of Sat1 is stripped by ram pressure from the hot diffuse halo gas only. 

When Sat1 and Sat2 enter the host galaxy, 26~\% and 29~\% of their total gas have a temperature below 10$^4$~K, respectively.
We confirm that the cold gas (T$<$10$^4$~K) of Sat1 is heated above 10$^4$~K after it is ram pressure stripped in the hot gas halo of the host galaxy.
Most of the stripped gas from Sat1 ($\sim$70~\%) is later blown out of the host galaxy by strong SN feedback driven winds originating in the disk of the host, while the rest cools and is later accreted onto the host's disk.
Sat2, meanwhile, loses 96~\% of its gas content when it crosses the host's disk. This form of gas loss enables the stripped gas to rapidly and efficiently cool and accrete onto the disk.
Indeed, we measure that 66~\% of the stripped gas from Sat2 is located on the disk and only 20~\% of the stripped gas is blown away from the host galaxy by $z=2.566$.

The satellites orbiting inside the host galaxy lose mass by tidal stripping.
Among the eight surviving satellites at $z=0$, five satellites that we label `ancient infallers' fall into the host galaxy at $z>2$ and the rest labelled `recent infallers' enter the host galaxy at $z<0.7$.
While orbiting inside the host galaxy, the `ancient infallers' lose more than 93~\% and 51~\% of their DM mass and stellar mass, respectively. 
On the other hand, the `recent infallers' lose only 78~\% and 2~\% of their DM and stellar mass, respectively.
Both subsamples show that the DM components of the satellites are preferentially lost as time goes by (as also seen in \citet{Han2018}), while the `recent infallers' show that the mass loss of the stellar component is delayed until the outer DM envelope is stripped \citep{Smith2016}. 
Even though the DM halo continues to be stripped over time, within their stellar disks they remain fully dominated by DM with $M_{s}/M_{t}<5\times10^{-3}$ and $M_{dyn} (<r_{h})>1.8\times10^{6}$~M$_{\odot}/h$, where $M_{dyn}(<r_{h})$ is the dynamical mass inside of the half light radius, $r_{h}$.

\section{Comparison with Observations}
\label{sec:observations}

\begin{figure}
\includegraphics[width=0.45\textwidth]{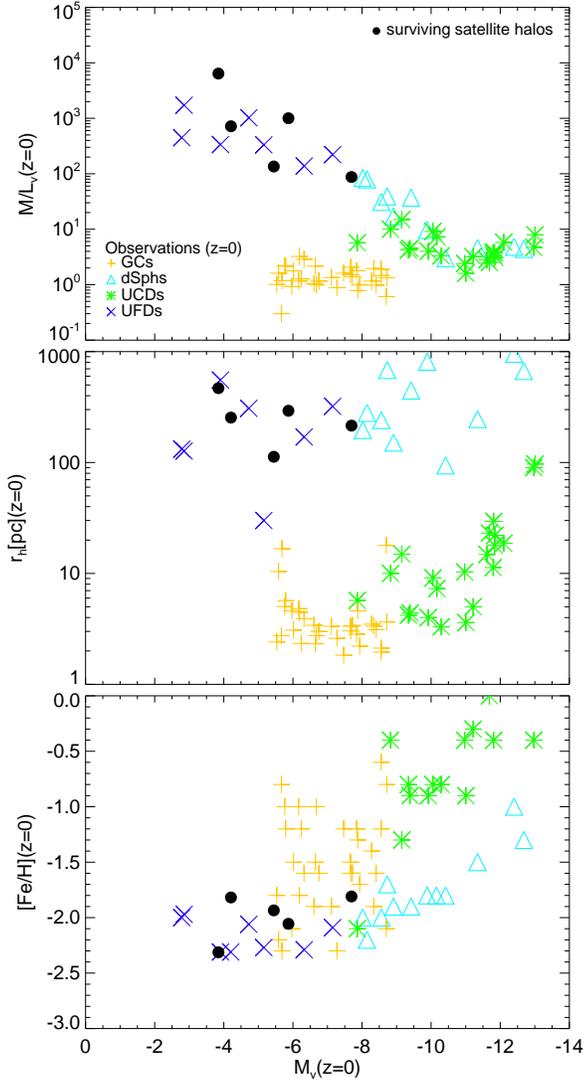}
\caption{Comparison between the observed satellites and the 5 surviving satellites in a space of $M/L_{V}-M_{V}$ (upper), $r_{h}-M_{V}$ (middle), and [Fe/H]$-M_{V}$ (lower). Observed GCs, dSphs, UCDs, and UFDs are indicated by yellow pluses, light blue triangles, green asterisks, and violet crosses, respectively.}
\label{fig:observation}
\end{figure}
Due to recent high-resolution photometric and spectroscopic observations, we now have more detailed information on satellite stellar objects with low luminosities, such as dSphs, UFDs, GCs, and UCDs \citep{Mateo1998,McConnachie2006,Evstigneeva2007,Simon2007,Mieske2008}.
In Fig. \ref{fig:observation}, we compare $M/L_{V}$, $r_{h}$, [Fe/H], and $M_{V}$ values for the surviving satellites with those of observed satellite stellar objects.
Since three of the 8 surviving satellites have no stellar components at $z=0$, only the remaining five are compared with the observed stellar objects at $z=0$.
We find that the values of $M_s/M_t(z_{in})$ of these five surviving satellites appears quite reasonable if we compare them to the extrapolated halo abundance matching results of \cite{Behroozi2013}.
The $M/L_{V}$ value for the surviving satellites is calculated with $M_{t}(r<r_{h})$ and $M_{s}(r<r_{h})$ at $z=0$.

The five surviving satellites have similar $M/L_{V}$, $r_{h}$, [Fe/H], and $M_{V}$  properties to those of UFDs.
Their mean stellar age is estimated to be $\sim$ 11~Gyr, which is comparable with the stellar age of observed UFDs ($\sim >$ 11.6~Gyr) \citep{Brown2014}. 

The UFD candidates have a size of $r_{h}$ = 0.11-0.47~kpc, which is quite similar to that the observed UFD, 0.1-0.55~kpc.

We do not find any star cluster candidates present in the mini-halos of our satellites. 
On the other hand, we find two star cluster candidates ($M_{s}=4\times10^{4}$ and $10^{5}$~M$_{\odot}/h$) that form in the gaseous disk of G1 at $z\sim3$. 
These star cluster candidates have very similar properties to observed GCs such as $M/L_{V}$ values as low as $\sim$ 1, and the metallicity, [Fe/H] $\sim$ −1.1. The objects are as large as the extended globular cluster ($r_{h}$ = 15-25~pc), but it is likely that the extended size is the result of the limited spatial resolution (12~pc/h) of our simulations.
Although both star cluster candidates do not survive until $z=0$, and have merged with their host by $z\sim0.7$, their presence supports a scenario in which GCs are generated within gas clumps in the host galaxy's gaseous disk \citep{Fall1985,Kravtsov2005,Kruijssen2015}, as opposed to forming directly with mini-halos.

\section{Summary and discussion}
\label{sec:summary}
Using high-resolution cosmological hydrodynamic simulations, we investigate how satellites around a dwarf galaxy form and evolve until $z=0$. 
For this, based on a parallel N-body/SPH code GADGET-3 \citep{Springel2005}, we implemented realistic astrophysical processes to describe baryonic structures of $M_{b} \geq 10^{4}$~M$_{\odot}/h$ \citep{Saitoh2006,Saitoh2009,Durier2012,Shin2014}. 
We performed three zoom simulations that target three different dwarf galaxies whose total mass at $z=0$ is $\sim 10^{10}$~M$_{\odot}/h$. 
Two of them evolve into dwarf disk galaxies, while the other evolves into a dwarf with more spherical morphology.

A total of 97 mini-halos of $M_{b} \geq 10^{4}$~M$_{\odot}/h$ form outside and then enter the host galaxy. 
Among them, 8 survive until the end of the runs, while the others are disrupted by the host galaxy’s tidal forces. 
The disrupted satellites contribute $\sim20~\%$ ($\sim5~\%$) of the host galaxy's DM (gas) mass growth, respectively. 
The characteristics of the mini-halos and satellites can be summarized as follows: 

1) The surviving satellites have similar properties to observed UFDs in terms of [Fe/H], $M/L_{V}$, $M_{V}$, $r_{h}$, and mean stellar age.
We find that two star cluster candidates form in the gaseous disk of the host galaxy. Although the candidates are disrupted by $z\sim0.7$, their presence supports a scenario in which globular clusters are formed within thermal instabilities or hydrodynamical shocks from collapsing gas clouds within the disk of their host galaxy \citep{Fall1985,Kravtsov2005,Kruijssen2015}.

2) In the case of the halos having stars, most of the gas is accreted at $z=3-5$. Although the peak of accretion of gas that will later contribute to star formation depends on the subsamples, most is accreted long after cosmic reionization. Therefore, the survival of a mini-halos gas to cosmic reionization is not an important factor controlling their stellar mass fractions.

3) More massive and higher concentration halos seem better able to cool hot gas that is later accreted, likely due to their ability to better compress the gas within their denser halos. We find that the halos in denser environments undergo more frequent merging events and more smooth accretion and thus grow their mass more rapidly. 

4) Tidal stripping of mini-halos by the host galaxy destroys 89 of 97 satellites. However, we find that the ram pressure is a crucial factor to quench their star formation, even though the host galaxy is a dwarf itself with a halo mass of only $M_{t}\leq 10^{10}$~M$_{\odot}/h$. 
The ram pressures can arise from motion of the mini-halos through both hot halo gas and also with the colder gaseous disk of the host galaxy.

Our results show that the properties of the satellites in the host galaxy are determined by both growth processes acting outside of the host galaxy as well as growth processes acting inside of the host galaxy.
In this paper, we investigated the three host galaxies only briefly because we choose to focus on the mini halos around the host galaxies instead.
However, in a future paper, we will investigate what drives the morphological differences we see between the host galaxies, and using a larger sample of dwarf galaxies.

\acknowledgments
This work was supported by a National Research Foundation grant funded by the Ministry of Science and ICT of Korea (NRF-2014R1A2A1A11052367). This work was also supported by the BK21 plus program through the National Research Foundation (NRF) funded by the Ministry of Education of Korea.

\appendix
\renewcommand\thefigure{A\arabic{figure}}
\setcounter{figure}{0}
\renewcommand\theequation{A\arabic{equation}}
\setcounter{equation}{0}

\cite{Okamoto2008b} performed cosmological hydrodynamic simulations to investigate which galaxies can accrete or hold onto their gas against the background UV field. 
They measure a characteristic mass, $M_c(z)$, directly from their simulations following the definition that it is the halo mass when the halo has a baryon mass fraction, $M_b/M_t$, of half the cosmic mean value. 
In their scenario, what matters is whether or not their halo mass is above or below the critical value of $M_c(z)$. 
However, both their halo mass and $M_c(z)$ evolves with time. 
For example, they find that $M_c(z)$ increases from $10^{7}$~M$_{\odot}/h$ right after reionization to $6.49\times10^{9}$~M$_{\odot}/h$ at $z=0$. 
They suggest that $M_c$ is determined by the temperature of the gas at the edge of the halo. Their simple model for gas accretion assumes the accreted gas at the edge of a halo has a density of one-third of the halo's virial gas density and this gas cannot accrete if it has a temperature higher than the halo's virial temperature,
\begin{equation} \label{eq:Tvir}
T_{vir}=\frac{1}{2}\frac{\mu m_p}{k_B}{V_c}^2,
\end{equation}
where, $\mu m_p$ is the average molecular mass and $k_B$ is the boltzmann constant, respectively, and the circular velocity, $V_c$, is measured at the virial radius of the halo.
They also found that background UV field prevents the gas from accreting to the halos with $M \leq M_c$ effectively. 

\begin{figure*}
\centering
\includegraphics[width=0.6\textwidth]{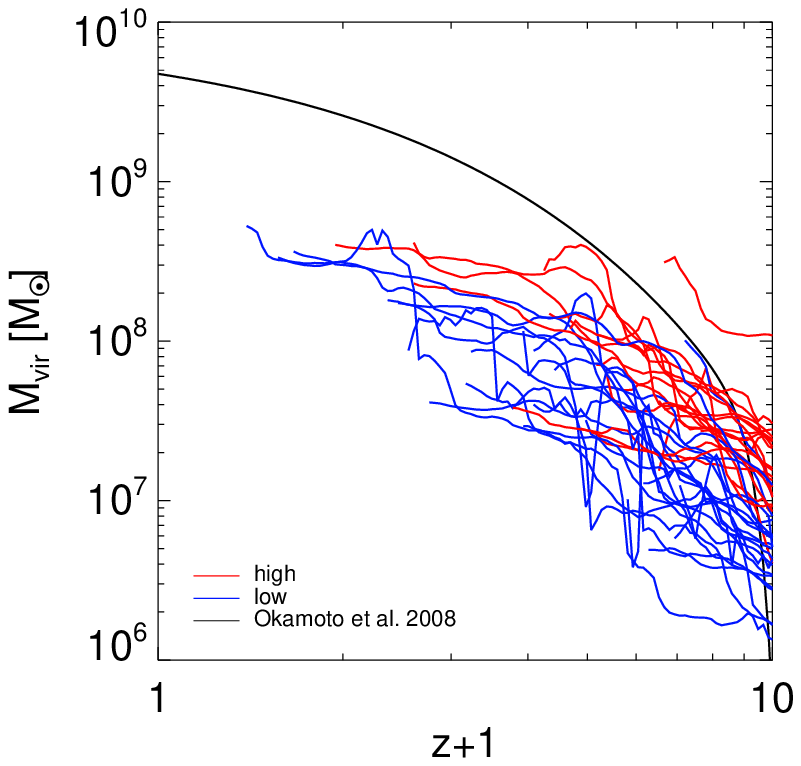}
\caption{The evolution of halo's mass and characteristic mass. Solid black, red, and blue lines represent the evolution of characteristic mass and virial mass of high and low stellar fraction subsample halos, respectively.}
\label{fig:mc}
\end{figure*}

In our simulations, we find that the high and low stellar fraction subsample halos are less massive than the characteristic mass in almost all of redshift (Fig. \ref{fig:mc}).
We find that 89~\% of these halos have their gas mass fraction reduced below half the cosmic mean value. The remaining 11~\% also lose at least 25~\% of their gas contents after reionization, but their early infall ($z_{in}>5.5$) and large self-shielded gas content prevents the gas fraction from dropping too low. 

The gas mass fraction of both subsamples continuously decreases, but the gas accretion is not fully halted.
To confirm whether the gas accretion model of \cite{Okamoto2008b} can be applied to the high and low stellar mass fraction subsamples, we investigate the temperature evolution of all the gas accreted after reionization that still remains when the halo first falls into the host galaxy.
We find, for both subsamples, 77~\% of total accreted gas has a temperature higher than the halo's virial temperature.
In particular, gas accretion peaks at $z=3-5$ and during this period, 97~\% of the accreted gas is hotter than the halo's virial temperature.
The fact that such hot gas can be accreted and does remain in the halos indicates that we cannot easily apply the gas accretion model of \cite{Okamoto2008b} to explain our subsamples.

One of the reasons why our simulations do not match the \cite{Okamoto2008b} results might be that our inclusion of a self-shielding recipe and cooling below 10$^4$ K which likely better enables the temperature of the bounded gas to continuously decrease, even if it is initially above the virial temperature of the halo. 
Indeed, we confirm that 39~\% of the gas hotter than the halo's virial temperature cools below the halo's virial temperature.
In addition, it is likely that the true temperature of the hot gas within a halo is not well approximated by the virial temperature equation given in Equation \ref{eq:Tvir}. 

%They also well predicted the characteristic mass evolution by gas accretion model which uses the merger tree of each halo and physical arguments about gas retention and accretion.
%In this gas accretion model, the gas at the virial radius of a halo cannot accrete to the halo if the gas has temperature higher than the halo's virial temperature. 
%The simple model assumes the gas at the edge of a halo has a density of one-third of the halo's virial gas density and the improved model determines the gas density based on the progenitor's baryon mass fraction.

%% This command is needed to show the entire author+affilation list when
%% the collaboration and author truncation commands are used.  It has to
%% go at the end of the manuscript.
%\allauthors

%% Include this line if you are using the \added, \replaced, \deleted
%% commands to see a summary list of all changes at the end of the article.
%\listofchanges

\end{document}